# Diamond-lattice photonic crystals assembled from DNA origami

Gregor Posnjak,[1,*] Xin Yin,[1] Paul Butler,[2,3] Oliver Bienek,[2,3] Mihir Dass,[1] Seungwoo Lee,[4,5] Ian D. Sharp,[2,3] and Tim Liedl[1,*]

[1]Faculty of Physics and CeNS, Ludwig-Maximilian-University Munich,
Schellingstraße 4, München, 80539, Bayern, Germany

[2]Walter Schottky Institute, Technical University of Munich,
Am Coulombwall 4, Garching bei München, 85748, Bayern, Germany

[3]Physics Department, TUM School of Natural Sciences, Technical University of Munich, Am Coulombwall 4, Garching bei München, 85748, Bayern, Germany

[4]Department of Integrative Energy Engineering (College of Engineering), KU-KIST Graduate School of Converging Science and Technology, Department of Biomicrosystem Technology, and KU Photonics Center, Korea University, 145 Anam-ro, Seongbuk-gu, Seoul, 02481, Republic of Korea

[5]Center for Opto-Electronic Materials and Devices, Post-Silicon Semiconductor Institute, Korea Institute of Science and Technology (KIST), 5 Hwarang-ro 14-gil, Seongbuk-gu, Seoul, 02792, Republic of Korea

[*]To whom correspondence should be addressed;
E-mail: gregor.posnjak@lmu.de, tim.liedl@physik.lmu.de.

**Colloidal self-assembly allows rational design of structures on the micrometer and submicrometer scale. One architecture that can generate complete 3D photonic band gaps is the diamond cubic lattice, which has remained difficult to realize at length scales comparable to the wavelength of visible or**

**ultraviolet light. Here, we demonstrate three-dimensional photonic crystals self-assembled from DNA origami that act as precisely programmable patchy colloids. Our DNA-based nanoscale tetrapods crystallize into a rod-connected diamond cubic lattice with a periodicity of 170 nm. This structure serves as a scaffold for atomic layer deposition of high refractive index materials such as $TiO_2$, yielding a tunable photonic band gap in the near-ultraviolet.**

## Introduction

Photonic crystals are materials with periodicity on length scales comparable to the wavelength of light, which together with the symmetry of their Brillouin zone leads to the formation of forbidden bands of energies for photons - the photonic band gap (PBG) (*1–3*). These optical analogs of semiconductors have remarkable properties such as omnidirectional reflections, lossless waveguiding, and suppressed emission. Such systems have been microfabricated in three dimensions (3D) for longer wavelengths (*4–6*), but their realization for visible and ultraviolet (UV) light has been limited because of the difficulty of their manufacturing (*7–9*). Moreover, most realizations of photonic crystals to date are based on surface-based microfabrication in two dimensions or possess only a limited number of layers forming the third dimension.

A promising approach for fabrication of 3D structures with submicrometer features is self-assembly of colloidal crystals (*10–12*). However, colloids typically form close-packed crystals with symmetries yielding relatively narrow photonic band gaps that open only in high-refractive index materials. Although this limitation can be circumvented by backfilling the voids within the colloidal crystals and selectively etching away the colloidal particles (*6, 8*), there remain limitations on achievable volume fill ratios and symmetries of the structures with this approach. Sequential synthesis and etching with different metals generate cage-like particles that form close-packed crystals of different symmetries with much larger voids than in the case of spherical colloids (*13*). However, the lossy metallic lattices are more suitable for different types of metamaterials rather than photonic crystals, and there are still considerable constraints in the types of structures that can be produced.

One of the most demanding geometries for colloidal crystallization is the diamond cubic lattice with its non-close-packed structure. The symmetry of the lattice requires that each building block has four neighbors in a tetrahedral configuration, which can be theoretically achieved with directional bonds (*14–16*), that is, with patchy particles that are challenging to realize experimentally. Generally, growth of such crystals is not favored because the hexagonal diamond lattice has the same free energy as the diamond cubic lattice, which causes the resulting material to be a mix of both structures (*15*).

In principle, this degeneracy can be broken either by introducing an orientation-dependent bonding potential on the binding patches (*14*) or by precisely controlling the connectivity of the four patches on each monomer (*17*). Both strategies have yet to be demonstrated experimentally. In addition to nanorobot-assisted assembly of diamond-type lattices on a lithographically patterned surface (*18*), several realizations of diamond-like structures have been achieved recently through assembly of gold nanoparticles (NPs) with DNA origami connectors (*19*), by directly connecting DNA origami frameworks into a diamond cubic lattice with 86 nm periodicity (*20*, *21*), and by forming a close-packed diamond crystal of micrometer-sized tetrahedral clusters of spheres (*22*). However, none of these realizations experimentally demonstrated a photonic band gap.

As a bottom-up approach, DNA origami (*23*) could enable unprecedented control over 3D colloidal assembly. This method combines the sequence-dependent selective binding of DNA, which has already been used in many colloidal crystal realizations (*12*), with particle design flexibility (*24*, *25*) and control over binding avidity and orientation (*26–30*). The selective binding of adenine (A) to thymine (T) and cytosine (C) to guanine (G) enables ~200 units of 20 to 50 nucleotide long staples to fold an ~8000 nucleotide long scaffold strand into predesigned shapes by formation of double-stranded helices (*23*). DNA origami has been used to assemble patchy particle-like objects that form non-close packed crystals through DNA-covered NP connectors (*19*), shape complementarity (*31*), or selective binding of short sequences of DNA (*20*).

Here, we demonstrate the use of DNA origami as patchy colloids with a programmed torsional potential on their binding patches to reliably orient neighboring monomers and assemble them into the diamond cubic lattice. We designed a DNA origami tetrapod in which each of its four arms served as a connecting patch to its neighbors. The pattern of binding extensions on the end surface of each arm provided the torsional binding potential favoring a 60° rotation between tetrapods (Fig. 1A). The tetrapods crystallized into a rod-connected diamond cubic lattice (Fig. 1B) that was predicted to show a wide, robust photonic band gap (*32*, *33*).

The 170-nm periodicity of our lattice led to a photonic band gap in the UV that would only open when the refractive index of the dielectric material in the lattice is higher than ~2 (*32*, *33*). Because of this requirement, we first silicified our structure to ensure mechanical stability during drying and then used atomic-layer deposition (ALD) (*34*) to grow varying thicknesses of high-refractive index materials on the surface of the lattice (Fig. 1C). We observed a strong reflection in the UV that was specific to the structure and tunable with the thickness of the high-refractive index cladding. With this approach, we realized a rod-connected diamond cubic lattice on the scale of a few hundred nanometers, as well as demonstrated a photonic effect with a DNA origami-based material.

## Crystal growth

We designed the DNA origami tetrapod structure in cadnano (*35*) (for design details, see fig. S1 and S2). Each tetrapod has four equivalent 35 nm long arms oriented at the tetrahedral angle of 109.5° (Fig. 1A). The cross-section of the arms is a 24-helix bundle (24hb) with an approximately circular cross section and a diameter of 15 nm. In the central part of the tetrapod, each arm was split into three 8-helix bundles (8hb) that were bent by ~70° by insertions and deletions of base pairs that induced over- and undertwisting of the DNA double helices (*36*) (fig. S2). Each of these bent 8hbs merged with two other 8hbs after the bend to form the neighboring arms of the tetrapod.

The double-stranded helices in each of the arms terminated at the same distance from the center of the tetrapod, resulting in a flat end surface. These end surfaces were modified with single-stranded DNA extensions of the staple strands to control the interaction between the monomers. With no extensions and hence blunt ends, the monomers aggregated uncontrollably and formed disordered networks even at temperatures near the melting temperature of the tetrapods (~ 55 °C (*37*), fig. S4). Therefore, we extended 18 of the end staples of each arm with 3 cytosines ($C_3$; Fig. 1A) to lower the temperature at which the monomers began to form lattices. Bonds between the tetrapods were formed by extending the staple strands on six of the helices on each end surface with a binding sequence TTTGGGAAGG (details on binding sequence design in Supplementary text and figs. S5 – S14). Although the tetrapods with the binding sequences have the correct symmetry to form the desired diamond cubic lattice, this was not a sufficient condition because they can form crystalline lattices by binding in two different conformations – the staggered conformation, where two neighboring tetrapods are rotated by 60° (fig. S15A), and the eclipsed conformation, where there is no rotation between two neighbors (fig. S15B). While the diamond cubic lattice is formed by tetrapods with only staggered conformations (Fig. 1B), in the hexagonal diamond lattice, each tetrapod is in the eclipsed configuration with one of its neighbors and in the staggered with the other three (fig. S15, C and D).

Crystallization of the diamond cubic lattice was further complicated by both types of diamond lattice having the same free energy, which would allow the tetrapods to crystallize as a mix of both lattices with many defects (*15*, *17*). To break this degeneracy, we placed the six binding sequences on each surface in a pattern with a threefold symmetry so that the binding between two neighboring tetrapods would be strongest when their terminal surfaces were rotated by 60°. All 12 binding extensions could form bonds between their guanines and the C$_3$ extensions on the other tetrapod (inset in Fig. 1A). Because all four arms of the tetrapod were equivalent, each pair of tetrapod neighbors was rotated by 60°, which led to the diamond cubic structure (Fig. 1B).

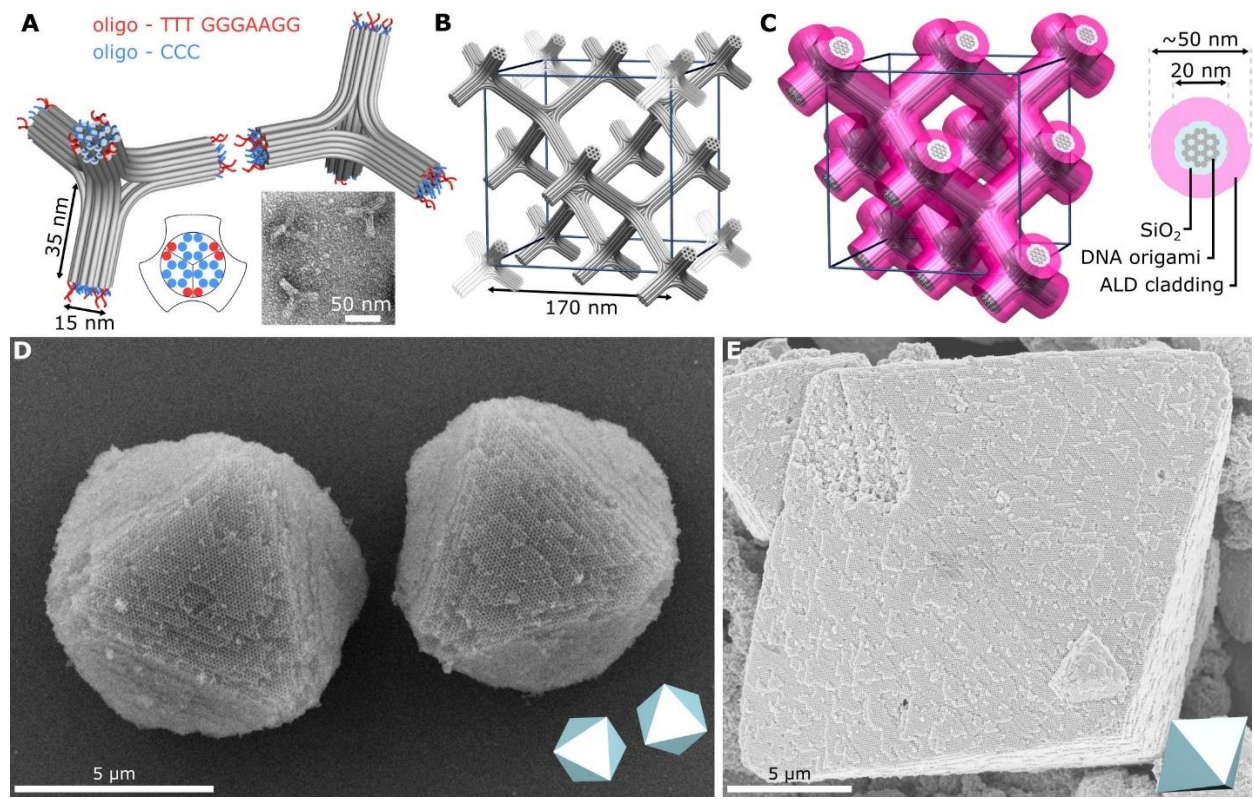

**Figure 1:** Design and growth of the diamond cubic crystals. (**A**) A model of two DNA origami tetrapods in the staggered configuration. Each gray cylinder in the model represents a double-stranded DNA helix. The binding sequence and the extensions of staples are shown in red and blue. The left inset shows the positions of extension on one of the end surfaces of the tetrapod. The right inset shows a TEM image of three tetrapods, with each showing only three arms, as the fourth arm is pointing out of the plane of the image. (**B**) Unit cell of the designed rod-connected diamond cubic lattice with a periodicity of 170 nm. (**C**) The unit cell, covered in layers of SiO$_2$ and ALD-grown high refractive index material. (**D**) Two DNA origami cubic diamond crystals, covered with a layer of SiO$_2$. (**E**) A 25 $\mu$m DNA origami diamond cubic crystal, covered with layers of SiO$_2$ and TiO$_2$. The octahedrons in the lower right corners of (D) and (E) illustrate the shape and orientation of the crystals.

In our initial experiments, the folded and purified monomers were slowly cooled with an 80 h ramp from 52 to 20 °C to form crystal lattices. This protocol already yielded crystals with approximately octahedral shape but rough surfaces and edges (fig. S16). After growth, all of the crystals were coated with a thin layer of silica ($SiO_2$) to increase their mechanical stability before drying (detailed protocol in Materials & Methods). After the silicification procedure, the thickness of our DNA origami structures was ~20 nm and the pores in the crystal were ~100 nm in diameter. The designed center-to-center distance of the tetrapods was 75 nm, which corresponded to a lattice periodicity of 170 nm.

The quality of the crystal structure was improved by incubating the crystal growth solution for 8 h at 40 °C after the first annealing ramp, pipetting away the top 70% of the supernatant, and then repeating the ramp. Under optimized conditions, diamond cubic single crystals exhibited regular octahedral crystal habits with all eight facets comprising equilateral triangles of {111} planes and dihedral angles over shared edges of 109.5°. Our crystal growth protocol resulted in most crystals having sizes between 5 and 10 $\mu$m (Fig. 1D and figs. S17 and S18), although some grew to as large as 25 $\mu$m (Fig. 1E). Based on the lattice periodicity of 170 nm and 8 tetrapods per unit cell, these crystals incorporated $10^5$ to $10^7$ tetrapods.

The {111} planes could have different appearances depending on the angle at which they are observed. When one of the faces of the octahedron was viewed along the {111} direction, it presented a hexagonal pattern in which the tetrapods of lower layers were also visible (Fig. 2A). If the same face was viewed at a slight tilt, connections between the tetrapods from the lower layers became visible (Fig. 2B). At an even larger tilt angle, the hexagonal grid appeared as two overlapping rectangular grids (Fig. 2C). The crystals we observed had surfaces with only a few steps between different {111} planes (Fig. 2D) and well-defined edges (Fig. 2E). However, the edges and especially the vertices of the octahedra were often ragged with missing tetrapods and facets exhibited irregular steps (figs. S19 and S20). The number of steps on the facets depended on the quality and speed of crystal growth. Additionally, the non-uniform edges and vertices could have been caused by mechanical forces during handling of the crystals, such as pipetting.

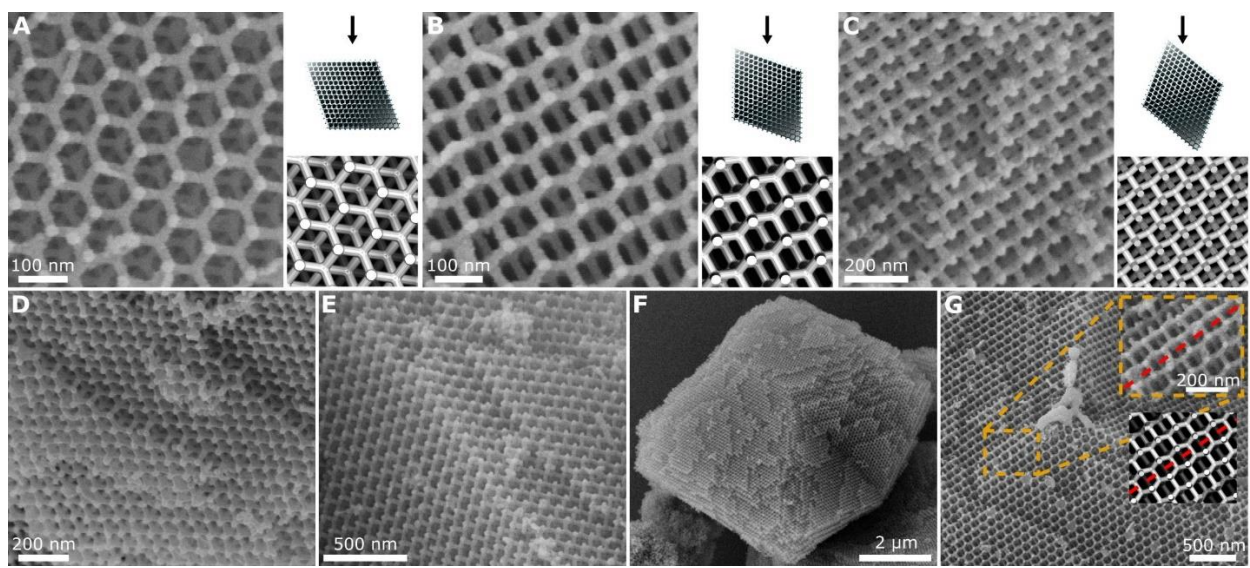

**Figure 2:** Structure of silica-coated diamond crystals. (**A to C**) SEM images of {111} facets of the crystal viewed at (A) normal orientation or tilted by (B) 20° and (C) 40° from the normal as shown in the upper insets. The lower insets show the appearance of a model of the crystal at the same orientation. (**D**) A zoomed-in view of steps of {111} planes on a crystal facet. (**E**) A well-defined edge between two crystal facets. (**F** and **G**) An example of a twinned crystal (F) where the mirror plane is a single plane of hexagonal diamond, indicated with the red dashed lines in the zoomed-in SEM image and the model shown in the inset of panel (G).

In rare cases we observed twinning of crystals, where the structure of the crystals was mirrored across a single plane (Fig. 2F). Closer inspection revealed that the mirror plane was an interface across which the tetrapods were not rotated by 60°, or in other words, a layer of hexagonal diamond (Fig. 2G). A hexagonal diamond defect in the otherwise diamond cubic structure appeared to be stable only if it formed an extended plane because we did not observe localized defects of this type. Such twinning planes appeared only in a few percent of the crystals, which suggests they were relatively unstable. Under non-optimal crystal growth conditions, where there were more disordered structures without well-defined habits, these hexagonal diamond defects were much more prevalent (fig. S21).

## Optical properties

To open a complete photonic band gap in a rod-connected or the more commonly discussed inverse diamond structure, the refractive index contrast between the air gaps and the dielectric material needs to be higher than ~2 (*32*, *33*). Because silica has a refractive index of ~1.5 we had to coat the silicified structure with a higher refractive index material. As a starting point we used MIT Photonic-Bands (MPB) (*38*) to simulate the photonic band structure of our crystal with an additional high-refractive index cladding of varying thickness. The spectral location and width of the photonic band gap change considerably with respect to the thickness of a high refractive index cladding and can be tuned across a relatively wide range of wavelengths, from 220 nm to 350 nm (fig. S22). Experimentally, we chose to coat our crystals with $TiO_2$ (titania) as this high refractive index material can be coated with ALD. The main advantage of ALD in our application is that the material grows layer by layer, guaranteeing a conformal coating on the structure with uniform thickness throughout the crystal.

Titania absorbs light below ∼350 nm, so its refractive index varied considerably across the range of wavelengths where we would expect a photonic band gap for our structure (fig. S23). Figure 3B shows a composite graph of calculated photonic band gaps for different thicknesses of $TiO_2$ coatings, where for each wavelength the relevant refractive index, measured experimentally using spectroscopic ellipsometry on a thin film reference sample, was used in the simulation (fig. S23). The relatively large variation of refractive index of $TiO_2$ resulted in a nonlinear variation of the PBG with varying volume fill ratio. At lower thicknesses of $TiO_2$ cladding, the simulations showed a very wide PBG that shifted to longer wavelengths and became progressively narrower with increasing thickness of $TiO_2$, influenced by the lower refractive index of titania at longer wavelengths. The measured extinction coefficient of titania (red curve in Fig. 3B) showed that a large portion of the possible photonic band gap was affected by relatively strong absorption, which could be mitigated by using a different ALD-compatible material, such as $Ta_2O_5$, which absorbs at shorter wavelengths than $TiO_2$ (fig. S24). However, it also has a lower refractive index, which leads to a narrower band gap with a more limited range of tunability of its central wavelength (fig. S25).

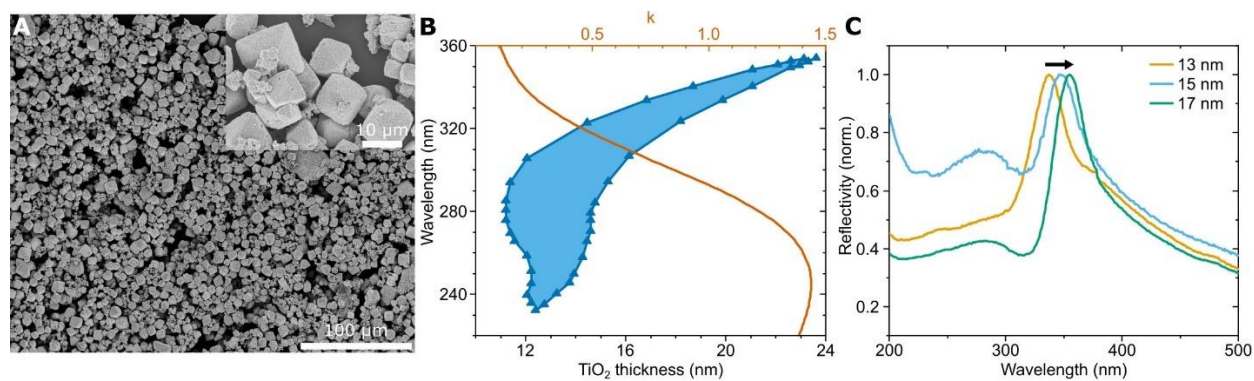

**Figure 3:** Optical properties of DNA origami photonic crystals. (**A**) Wide field and zoomed-in (inset) SEM images of a deposited film of silicified DNA origami diamond crystals, covered with a 13 nm layer of titania ($TiO_2$). (**B**) Calculated photonic band gaps for different thicknesses of $TiO_2$ cladding on the diamond cubic structure and the measured extinction coefficient (k) of ALD-grown $TiO_2$. The photonic band gap is calculated using measured values of the refractive index of ALD-deposited films. (**C**) Measured reflection spectra of DNA origami diamond crystal films with different thicknesses of titania coating.

Films of diamond crystals coated with titania (Fig. 3A) were characterized by optical reflectivity measurements. Reflectivity measurements of our samples showed a strong reflection for wavelengths approximately corresponding to the calculated PBGs (Fig. 3C). With increasing thickness of $TiO_2$ the band of reflected wavelengths redshifted, as predicted in the simulations. To verify that this reflection was a consequence of the structure of our crystals, we prepared three additional types of structures – non-crystalline assemblies of the tetrapods with center-to-center distance between neighboring tetrapods equal to 75 nm (the same as in our crystalline samples; fig. S26, A and B), non-crystalline assemblies with extension bundles between the tetrapods (center-to-center distance of 180 nm; fig. S26, C and D), and random aggregates of tetrapods without consistent local structure, but with grain sizes roughly comparable to sizes of the crystals (fig. S26, E and F). Importantly, fig. S26, G and H, show that for all three reference samples the reflection appeared at longer wavelengths than for the crystals, and its position and spectral shape were independent of the structure (fig. S26G) and the thickness of titania coating (fig. S26H and Supplementary text).

This spectral feature in non-crystalline samples was not related to a photonic band gap. We attribute the spectral shape of these reflections to a combination of Mie scattering from the micrometer-scale roughness of the samples and absorption of light by defect centers in the ALD deposited $TiO_2$ (*39*, *40*). Neither of these effects was accounted for in the simulations of the PBG. By contrast, the spectral reflectivity of the diamond cubic crystal was consistent with optical modeling and could be attributed to the opening of a photonic band gap. Ultimately, single-crystal angle-dependent spectroscopic measurements (*41*) or observation of depletion of the optical density of states (*1*, *42*, *43*) would conclusively prove the formation of a complete photonic band gap.

**Discussion**

Although the spectral response of our crystals was limited by the absorption of titania, we demonstrated the use of diamond cubic structures as templates for photonic crystals. Their potential could be maximized by using a more transparent material for the high refractive index cladding and by tuning their lattice constant by either co-crystallization of multiple building blocks (*44*) or the use of multiple scaffolds (*32*, *45*) to increase the size of the tetrapods. The sequence-specific binding of DNA origami-based building blocks could be used for self-assembly of precisely controlled, scalable (details in Supplementary text) and highly customizable architectures needed for realizations of different three-dimensional photonic crystals and topological photonic concepts. This programmable bottom-up manufacturing approach could lead to a vastly extended design space for manufacturing of optical circuits (*46*).

# References

1. E. Yablonovitch, Inhibited Spontaneous Emission in Solid-State Physics and Electronics. *Phys. Rev. Lett.* **58**, 2059–2062 (1987).

2. S. John, Strong localization of photons in certain disordered dielectric superlattices. *Phys. Rev. Lett.* **58**, 2486–2489 (1987).

3. E. Yablonovitch, T. J. Gmitter, Photonic band structure: The face-centered-cubic case. *Phys. Rev. Lett.* **63**, 1950–1953 (1989).

4. S. Y. Lin, J. G. Fleming, D. L. Hetherington, B. K. Smith, R. Biswas, K. M. Ho, M. M. Sigalas, W. Zubrzycki, S. R. Kurtz, J. Bur, A three-dimensional photonic crystal operating at infrared wavelengths. *Nature* **394**, 251–253 (1998).

5. S. Noda, K. Tomoda, N. Yamamoto, A. Chutinan, Full Three-Dimensional Photonic Bandgap Crystals at Near-Infrared Wavelengths. *Science* **289**, 604–606 (2000).

6. A. Blanco, E. Chomski, S. Grabtchak, M. Ibisate, S. John, S. W. Leonard, C. Lopez, F. Meseguer, H. Miguez, J. P. Mondia, G. A. Ozin, O. Toader, H. M. van Driel, Large-scale synthesis of a silicon photonic crystal with a complete three-dimensional bandgap near 1.5 micrometres. *Nature* **405**, 437–440 (2000).

7. M. Campbell, D. N. Sharp, M. T. Harrison, R. G. Denning, A. J. Turberfield, Fabrication of photonic crystals for the visible spectrum by holographic lithography. *Nature* **404**, 53–56 (2000).

8. Y. A. Vlasov, X.-Z. Bo, J. C. Sturm, D. J. Norris, On-chip natural assembly of silicon photonic bandgap crystals. *Nature* **414**, 289–293 (2001).

9. G. Subramania, Y.-J. Lee, A. J. Fischer, D. D. Koleske, Log-Pile TiO2 Photonic Crystal for Light Control at Near-UV and Visible Wavelengths. *Adv. Mater.* **22**, 487–491 (2010).

10. R. J. Macfarlane, B. Lee, M. R. Jones, N. Harris, G. C. Schatz, C. A. Mirkin, Nanoparticle Superlattice Engineering with DNA. *Science* **334**, 204–208 (2011).

11. M. R. Jones, N. C. Seeman, C. A. Mirkin, Programmable materials and the nature of the DNA bond. *Science* **347**, 1260901 (2015).

12. C. A. Mirkin, S. H. Petrosko, Inspired Beyond Nature: Three Decades of Spherical Nucleic Acids and Colloidal Crystal Engineering with DNA. *ACS Nano* **17**, 16291–16307 (2023).

13. Y. Li, W. Zhou, I. Tanriover, W. Hadibrata, B. E. Partridge, H. Lin, X. Hu, B. Lee, J. Liu, V. P. Dravid, K. Aydin, C. A. Mirkin, Open-channel metal particle superlattices. *Nature* **611**, 695–701 (2022).

14. Zhang, A. S. Keys, T. Chen, S. C. Glotzer, Self-Assembly of Patchy Particles into Diamond Structures through Molecular Mimicry. *Langmuir* **21**, 11547–11551 (2005).

15. F. Romano, E. Sanz, F. Sciortino, Crystallization of tetrahedral patchy particles in silico. *J. Chem. Phys.* **134**, 174502 (2011).

16. Y. Wang, Y. Wang, D. R. Breed, V. N. Manoharan, L. Feng, A. D. Hollingsworth, M. Weck, D. J. Pine, Colloids with valence and specific directional bonding. *Nature* **491**, 51–55 (2012).

17. L. Rovigatti, J. Russo, F. Romano, M. Matthies, L. Kroc, P. Šulc, A simple solution to the problem of self-assembling cubic diamond crystals. *Nanoscale* **14**, 14268–14275 (2022).

18. F. García-Santamaría, H. t. Miyazaki, A. Urquía, M. Ibisate, M. Belmonte, N. Shinya, F. Meseguer, C. López, Nanorobotic Manipulation of Microspheres for On-Chip Diamond Architectures. *Adv. Mater.* **14**, 1144–1147 (2002).

19. W. Liu, M. Tagawa, H. L. Xin, T. Wang, H. Emamy, H. Li, K. G. Yager, F. W. Starr, A. V. Tkachenko, O. Gang, Diamond family of nanoparticle superlattices. *Science* **351**, 582–586 (2016).

20. Y. Tian, J. R. Lhermitte, L. Bai, T. Vo, H. L. Xin, H. Li, R. Li, M. Fukuto, K. G. Yager, J. S. Kahn, Y. Xiong, B. Minevich, S. K. Kumar, O. Gang, Ordered three-dimensional nanomaterials using DNA-prescribed and valence-controlled material voxels. *Nat. Mater.* **19**, 789–796 (2020).

21. A. Michelson, B. Minevich, H. Emamy, X. Huang, Y. S. Chu, H. Yan, O. Gang, Three-dimensional visualization of nanoparticle lattices and multimaterial frameworks. *Science* **376**, 203–207 (2022).

22. M. He, J. P. Gales, É. Ducrot, Z. Gong, G.-R. Yi, S. Sacanna, D. J. Pine, Colloidal diamond. *Nature* **585**, 524–529 (2020).

23. P. W. K. Rothemund, Folding DNA to create nanoscale shapes and patterns. *Nature* **440**, 297–302 (2006).

24. S. M. Douglas, H. Dietz, T. Liedl, B. Högberg, F. Graf, W. M. Shih, Self-assembly of DNA into nanoscale three-dimensional shapes. *Nature* **459**, 414–418 (2009).

25. D. Han, S. Pal, J. Nangreave, Z. Deng, Y. Liu, H. Yan, DNA origami with complex curvatures in three-dimensional space. *Science* **332**, 342–346 (2011).

26. J. Zheng, J. J. Birktoft, Y. Chen, T. Wang, R. Sha, P. E. Constantinou, S. L. Ginell, C. Mao, N. C. Seeman, From molecular to macroscopic via the rational design of a self-assembled 3D DNA crystal. *Nature* **461**, 74–77 (2009).

27. T. Gerling, K. F. Wagenbauer, A. M. Neuner, H. Dietz, Dynamic DNA devices and assemblies formed by shape-complementary, non–base pairing 3D components. *Science* **347**, 1446–1452 (2015).

28. D. Minev, C. M. Wintersinger, A. Ershova, W. M. Shih, Robust nucleation control via crisscross polymerization of highly coordinated DNA slats. *Nat. Commun.* **12**, 1741 (2021).

29. M. Ji, Z. Zhou, W. Cao, N. Ma, W. Xu, Y. Tian, A universal way to enrich the nanoparticle lattices with polychrome DNA origami "homologs." *Sci. Adv.* **8**, eadc9755 (2022).

30. C. M. Wintersinger, D. Minev, A. Ershova, H. M. Sasaki, G. Gowri, J. F. Berengut, F. E. Corea-Dilbert, P. Yin, W. M. Shih, Multi-micron crisscross structures grown from DNA-origami slats. *Nat. Nanotechnol.* **18**, 281–289 (2023).

31. T. Zhang, C. Hartl, K. Frank, A. Heuer-Jungemann, S. Fischer, P. C. Nickels, B. Nickel, T. Liedl, 3D DNA origami crystals. *Adv. Mater.* **30**, 1800273 (2018).

32. S. H. Park, H. Park, K. Hur, S. Lee, Design of DNA Origami Diamond Photonic Crystals. *ACS Appl. Bio Mater.* **3**, 747–756 (2020).

33. R. K. Cersonsky, J. Antonaglia, B. D. Dice, S. C. Glotzer, The diversity of three-dimensional photonic crystals. *Nat. Commun.* **12**, 2543 (2021).

34. R. W. Johnson, A. Hultqvist, S. F. Bent, A brief review of atomic layer deposition: from fundamentals to applications. *Mater. Today* **17**, 236–246 (2014).

35. S. M. Douglas, A. H. Marblestone, S. Teerapittayanon, A. Vazquez, G. M. Church, W. M. Shih, Rapid prototyping of 3D DNA-origami shapes with caDNAno. *Nucleic Acids Res.* **37**, 5001–5006 (2009).

36. H. Dietz, S. M. Douglas, W. M. Shih, Folding DNA into Twisted and Curved Nanoscale Shapes. *Science* **325**, 725–730 (2009).

37. H. Ijäs, T. Liedl, V. Linko, G. Posnjak, A label-free light-scattering method to resolve assembly and disassembly of DNA nanostructures. *Biophys. J.* **121**, 4800–4809 (2022).

38. S. G. Johnson, J. D. Joannopoulos, Block-iterative frequency-domain methods for Maxwell's equations in a planewave basis. *Opt. Express* **8**, 173–190 (2001).

39. H. Hu, T. Weber, O. Bienek, A. Wester, L. Hüttenhofer, I. D. Sharp, S. A. Maier, A. Tittl, E. Cortés, Catalytic Metasurfaces Empowered by Bound States in the Continuum. *ACS Nano* **16**, 13057–13068 (2022).

40. O. Bienek, B. Fuchs, M. Kuhl, T. Rieth, J. Kühne, L. I. Wagner, L. M. Todenhagen, L. Wolz, A. Henning, I. D. Sharp, Engineering Defects and Interfaces of Atomic Layer-Deposited TiOx Protective Coatings for Efficient III−V Semiconductor Photocathodes. *ACS Photonics*, doi: https://doi.org/10.1021/acsphotonics.3c00818 (2023).

41. E. Palacios-Lidón, A. Blanco, M. Ibisate, F. Meseguer, C. López, J. Sánchez-Dehesa, Optical study of the full photonic band gap in silicon inverse opals. *Appl. Phys. Lett.* **81**, 4925–4927 (2002).

42. P. Lodahl, A. Floris van Driel, I. S. Nikolaev, A. Irman, K. Overgaag, D. Vanmaekelbergh, W. L. Vos, Controlling the dynamics of spontaneous emission from quantum dots by photonic crystals. *Nature* **430**, 654–657 (2004).

43. M. Fujita, S. Takahashi, Y. Tanaka, T. Asano, S. Noda, Simultaneous Inhibition and Redistribution of Spontaneous Light Emission in Photonic Crystals. *Science* **308**, 1296–1298 (2005).

44. N. Ma, L. Dai, Z. Chen, M. Ji, Y. Wang, Y. Tian, Environment-Resistant DNA Origami Crystals Bridged by Rigid DNA Rods with Adjustable Unit Cells. *Nano Lett.* **21**, 3581–3587 (2021).

45. F. A. S. Engelhardt, F. Praetorius, C. H. Wachauf, G. Brüggenthies, F. Kohler, B. Kick, K. L. Kadletz, P. N. Pham, K. L. Behler, T. Gerling, H. Dietz, Custom-Size, Functional, and Durable DNA Origami with Design-Specific Scaffolds. *ACS Nano* **13**, 5015–5027 (2019).

46. A. W. Elshaari, W. Pernice, K. Srinivasan, O. Benson, V. Zwiller, Hybrid integrated quantum photonic circuits. *Nat. Photonics* **14**, 285–298 (2020).

47. E. Stahl, T. G. Martin, F. Praetorius, H. Dietz, Facile and Scalable Preparation of Pure and Dense DNA Origami Solutions. *Angew. Chem. Int. Ed.* **53**, 12735–12740 (2014).

48. L. Nguyen, M. Döblinger, T. Liedl, A. Heuer-Jungemann, DNA-Origami-Templated Silica Growth by Sol–Gel Chemistry. *Angew. Chem. Int. Ed.* **58**, 912–916 (2019).

49. P. W. Majewski, A. Michelson, M. A. L. Cordeiro, C. Tian, C. Ma, K. Kisslinger, Y. Tian, W. Liu, E. A. Stach, K. G. Yager, O. Gang, Resilient three-dimensional ordered architectures assembled from nanoparticles by DNA. *Sci. Adv.* **7**, eabf0617 (2021).

50. D.-N. Kim, F. Kilchherr, H. Dietz, M. Bathe, Quantitative prediction of 3D solution shape and flexibility of nucleic acid nanostructures. *Nucleic Acids Res.* **40**, 2862–2868 (2012).

51. K. Edagawa, S. Kanoko, M. Notomi, Photonic Amorphous Diamond Structure with a 3D Photonic Band Gap. *Phys. Rev. Lett.* **100**, 013901 (2008).

52. H. Yin, B. Dong, X. Liu, T. Zhan, L. Shi, J. Zi, E. Yablonovitch, Amorphous diamond-structured photonic crystal in the feather barbs of the scarlet macaw. *Proc. Natl. Acad. Sci. U.S.A.* **109**, 10798–10801 (2012).

53. F. Praetorius, B. Kick, K. L. Behler, M. N. Honemann, D. Weuster-Botz, H. Dietz, Biotechnological mass production of DNA origami. *Nature* **552**, 84–87 (2017).

# Acknowledgments

We thank E. V. Sturm for valuable discussions and M. Herrán and C. Fan for the help with UV-Vis measurements. We thank T. Rieth for developing the $Ta_2O_5$ ALD protocol. **Funding:** G.P., X.Y. and T.L. acknowledge funding from the ERC consolidator grant "DNA Funs" (Project ID: 818635). M.D. acknowledges financial support from the Federal Ministry of Education and Research (BMBF) and the Free State of Bavaria under the Excellence Strategy of the Federal

Government and the Länder through the ONE MUNICH Projects Munich Multiscale BioFabrication, and Enabling Quantum Communication and Imaging Applications. P.B., O.B. and I.D.S. acknowledge financial support from TUM.Solar in the context of the Bavarian Collaborative Research Project Solar Technologies Go Hybrid (SolTech). T.L., P.B., O.B. and I.D.S. further acknowledge support from the Deutsche Forschungsgemeinschaft (DFG, German Research Foundation) through the cluster of excellence e-conversion EXC 2089/1-390776260. S.L. acknowledges funding from the National Research Foundation of Korea (NRF-2022M3H4A1A02074314 and NRF-RS-2023-00272363). **Author contributions:** T.L. and S.L. initiated the research. T.L. and G.P. designed and supervised the experiments. G.P. designed the tetrapod structure. G.P. and X.Y. conducted the crystallization and optical experiments with assistance from M.D.. X.Y. performed the PBG simulations supervised by G.P.  P.B. and O.B. performed the ALD coating. I.D.S. supervised the ALD experiments. G.P. and T.L. wrote the manuscript with inputs from all authors. **Competing interests:** None declared. **Data and materials availability:** All data needed to evaluate the conclusions in the paper are present in the paper or the Supplementary Materials.

# Supplementary materials

Materials and Methods

Supplementary Text available on DOI: 10.1126/science.adl2733

Figs. S1 to S26

Tables S1 and S2

References (47 - 53)